\pdfoutput=1
      [1999/12/01 v1.4c Il Nuovo Cimento]
\documentclass{cimento}

\usepackage{graphicx}  
\title{Top Quark Spin Correlations - Theory}
\author{Stephen~J.~Parke\from{ins:x}\thanks{Presented at TOP2011 in Sant Feliu de Guixols, Spain, Sept 25-30, 2011.}}
\instlist{\inst{ins:x} Theoretical Physics Dept., Fermi National Accelerator Laboratory, Batavia IL  60510, USA} 
\begin{document}

\maketitle

\begin{abstract}
The theoretical aspects of spin correlations in top quark pair production are briefly reviewed.
\end{abstract}

The top quark decay width ($ G_F m_t^3 \sim$  1 GeV) is much larger than the QCD hadronization scale ($\Lambda_{QCD} \sim $ 0.1 GeV) and much larger than the spin decorrelation scale\\ ($\Lambda_{QCD}^2/m_t \sim$ 0.1 MeV). Therefore, spin correlations in top quark pair production are reflected in angular correlations of the
decay products, see \cite{Mahlon:1995zn} and \cite{Stelzer:1995gc}.

\section{Top Quark Pair Production}
\subsection{Quark-antiquark annihilation or unlike helicity gluon fusion}
For top quark pair production via quark-antiquark annihilation or unlike helicity gluon fusion, there exists a spin axis such that the top quarks are produced in only
the up-down or down-up configuration, i.e. parallel, since the spin axes are back to back,
\begin{eqnarray}
q_L \bar{q}_R, ~q_L \bar{q}_R, ~g_L g_R, ~g_R g_L  \rightarrow  t_U \bar{t}_D + t_D \bar{t}_U.
\end{eqnarray}
No combinations $ t_U \bar{t}_U$ or $t_D \bar{t}_D$ are produced, see Fig. 1. This spin basis is known as the Off-Diagonal basis, see \cite{Parke:1996pr} and \cite{Mahlon:2010gw},
and the spin axis makes an angle $\Omega$ wrt to the top quark momentum direction in the ZM frame. This angle is given by
\begin{eqnarray}
\tan \Omega  = (1-\beta^2) \tan \theta = \frac{1}{\gamma^2} \tan \theta.
\end{eqnarray}
Where the speed and the scattering angle of the top quark are given by $\beta$ and $\theta$, respectively.  Note, at threshold, $\Omega=\theta$ at the spin axis is aligned along the beamline, whereas at ultra-high energies, $\Omega=0$ and the spin axis is aligned along the direction of motion of the top quark (ZMF helicity).
\begin{figure}
\begin{center}
\includegraphics[scale=0.4]{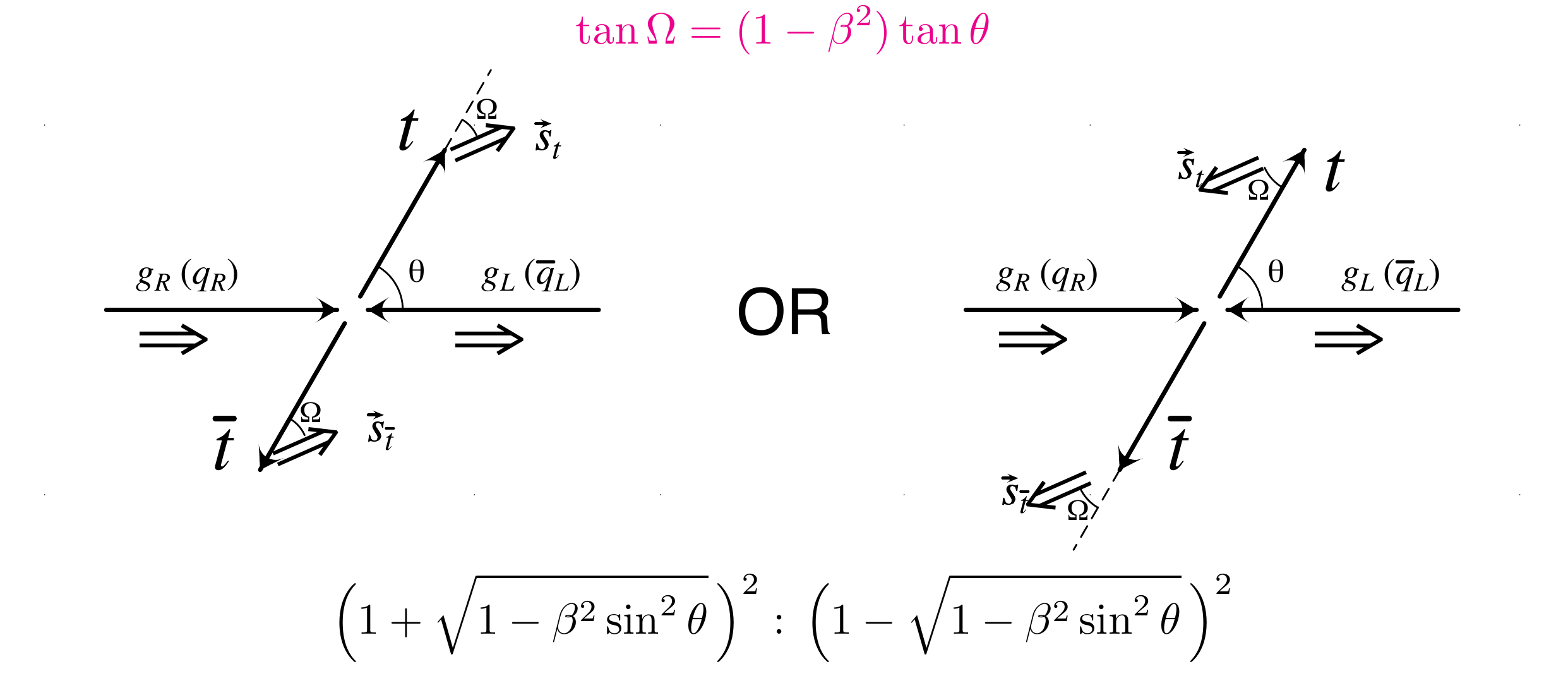}     
\end{center}
\caption{Spin correlations for the production of top quark pairs via unlike helicity gluon fusion or quark-antiquark annihilation.}
\end{figure}

\subsection{Like helicity gluon fusion}
For top quark pair production via like helicity gluon fusion in the helicity basis, the top quarks are produced in only
the left-left or right-right configuration, i.e. anti-parallel,
\begin{eqnarray}
g_L g_L, ~g_R g_R  \rightarrow  t_L \bar{t}_L + t_R \bar{t}_R.
\end{eqnarray}
No combinations $ t_L \bar{t}_R$ or $t_R \bar{t}_L$ are produced in this process, see Fig. 2.
\begin{figure}
\begin{center}
\includegraphics[scale=0.4]{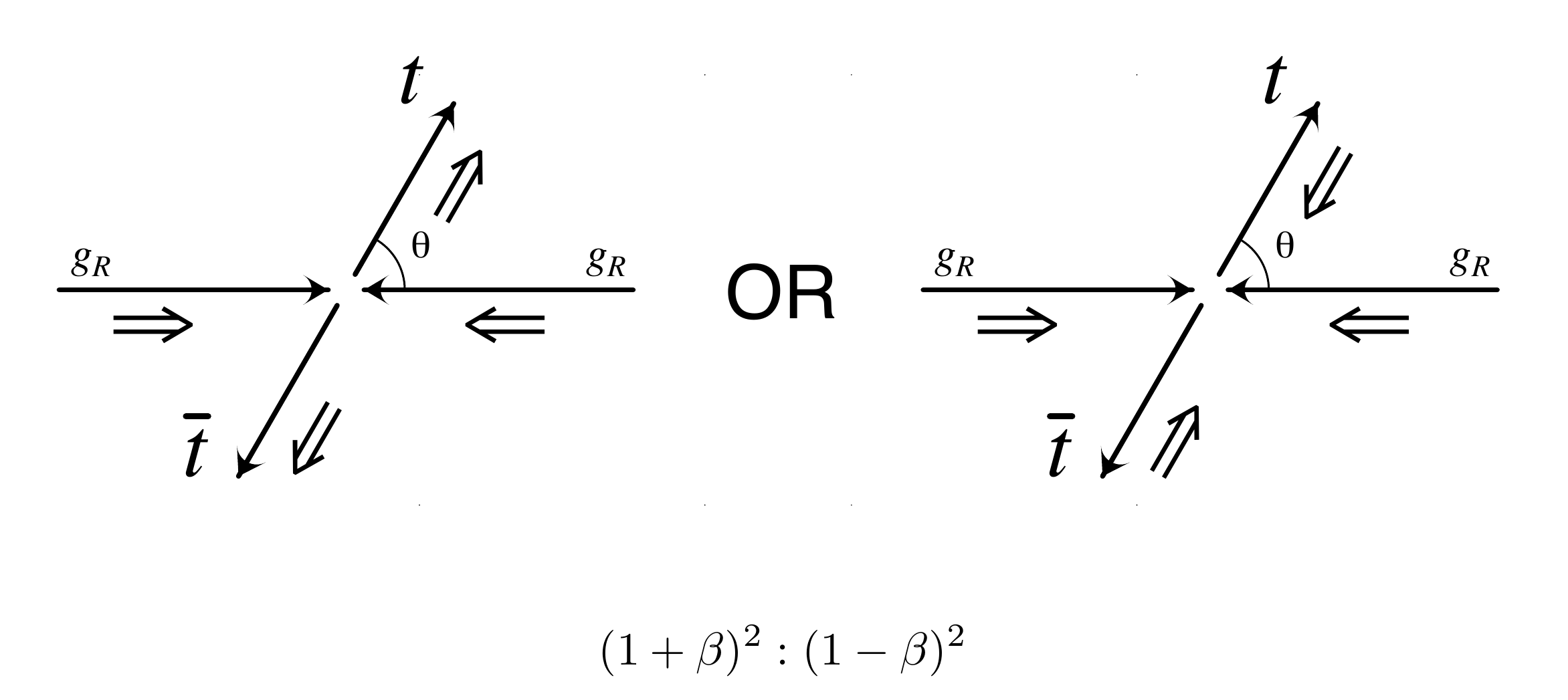}     
\end{center}
\caption{Spin correlations for the production of top quark pairs via like helicity gluon fusion.}
\end{figure}

\section{Polarized top quark decay}
The decay products from a polarized top quark have their moment vectors correlated with the top quark spin axis as follows:
\begin{eqnarray}
\hspace*{1cm} \frac{1}{\Gamma_T} \frac{d \Gamma}{d \cos \chi_i} 
= (1+\alpha_i \cos \chi_i)/2, \quad {\rm where} \quad \alpha_i=\left\{ \begin{array}{lll}
+1.0 &(+0.998)   ~~&\hbox{$l^+$ }\\
+1.0 &(+0.966)  &\hbox{$\bar{d}$-quark}\\
-0.31 &(-0.314)  &\hbox{$\bar\nu$ } \\
-0.31 &(-0.317)   &\hbox{$u$-quark} \\
-0.41 &(-0.393)   &\hbox{$b$-quark}
  \end{array} \right.
\end{eqnarray}
at LO with the NLO results, see   \cite{Bernreuther:2001rq}, in parenthesis. $\chi_i$ is the angle between the decay product and the spin axis in the top quark rest frame.

\section{Understanding the Full Correlated Matrix Element Squared}
In top quark rest frame, denote the polar and azimuthal angles of the charged lepton from top quark decay wrt to the top spin axis by $\chi$ and $\phi$ respectively. Whereas for the anti-top quark these angles wrt to the anti-top quark spin axis by $\bar{\chi}$ and $\bar{\phi}$ respectively.  $\phi$ and $\bar{\phi}$ are determined wrt to the scattering plane. Then for quark-antiquark annihilation, the correlated part of the full matrix element squared, apart from over all factor $(b\cdot \nu)(\bar{b}\cdot \bar{\nu})$, is given by
 \begin{eqnarray}
|{\cal A}|_{LR}^2 + |{\cal A}|_{RL}^2 & \sim & ~[(2-\beta^2s^2_\theta) {(1+c_\chi c_{\bar{\chi}})} ~{+~\beta^2 s^2_\theta  s_\chi s_{\bar{\chi}} \cos(\phi+\bar{\phi}) }]
\end{eqnarray}
in the Off-Diagonal bases. Also for unlike helicity gluon fusion 
 \begin{eqnarray}
~~~~|{\cal A}|_{LR}^2 + |{\cal A}|_{RL}^2 & \sim &  \frac{(7+9 \beta^2 c^2_\theta)}{(1-\beta^2 c^2_\theta)^2}~\beta^2 s^2_\theta~[(2-\beta^2s^2_\theta) {(1+c_\chi c_{\bar{\chi}})} ~{+~\beta^2 s^2_\theta  s_\chi s_{\bar{\chi}} \cos(\phi+\bar{\phi}) }]
\end{eqnarray}
again in the Off-Diagonal bases. Notice that the part of these two expressions between $[\cdots]$ are identical, so that these two expressions only differ by an overall factor.

Whereas for  like helicity gluon fusion in the helicity bases, we have
 \begin{eqnarray}
~~~~~|{\cal A}|_{LL}^2 + |{\cal A}|_{RR}^2 & \sim  & \frac{(7+9 \beta^2 c^2_\theta)}{(1-\beta^2 c^2_\theta)^2}~(1-\beta^2)[(1+\beta^2)(1-c_\chi c_{\bar{\chi}})~{-(1-\beta^2)s_\chi s_{\bar{\chi}} \cos(\phi-\bar{\phi})}]
\end{eqnarray}
 in the helicity basis.
So the full correlation is determined by the angles of the charge leptons (or down type quarks) wrt the spin axis and momentum conservation.  The term proportional to 
$\cos(\phi- \bar{\phi})$ ($\cos(\phi + \bar{\phi})$) is the interference terms between the $t_L \bar{t}_L$ \& $t_R \bar{t}_R$ ($t_U \bar{t}_D$ \& $t_D \bar{t}_U$ ) for like helicity gluon fusion (unlike helicity gluon fusion or quark-antiquark annihilation). Eqn (5), (6) and (7) can be easily derived from Eqn (B1), (47) and (48) of ref. \cite{Mahlon:2010gw} respectively.

For the uncorrelated case, just set $c_\chi c_{\bar{\chi}}$ and $\cos(\phi\pm \bar{\phi})$ equal to their uncorrelated average values, zero!
The expressions for arbitrary spin axis can be found in the slides presented at the workshop.

\section{Observables}
\subsection{The Effects of Spin Correlations}
The dominant effect of the spin correlations is to correlated the angles of the decay products between the top quark and antitop quark,
i.e. between $\chi_i$ and $\bar{\chi}_{\bar{i}}$. This correlation is given by
\begin{eqnarray}
\frac{1}{\sigma_T} \frac{d^2 \sigma}{d \cos \chi_i  ~d \cos \bar{\chi}_{\bar{i} }  }=  \frac{1}{4} (1+C_{t\bar{t}} ~\alpha_i \alpha_{\bar{i}} ~ \cos \chi_i \cos \bar{\chi}_{\bar{i}}) 
\end{eqnarray}
where
\vspace{-0.5cm}
\begin{eqnarray}
C_{t\bar{t}} \equiv \frac{\sigma_{\uparrow \uparrow}+\sigma_{\downarrow \downarrow}-\sigma_{\uparrow \downarrow}-\sigma_{\downarrow \uparrow}}
{\sigma_{\uparrow \uparrow}+\sigma_{\downarrow \downarrow}+\sigma_{\uparrow \downarrow}+\sigma_{\downarrow \uparrow}}
=\left\{ \begin{array}{lll}
-0.456 & (-0.389) & {\rm Helicity ~at ~Tevatron}\\
+0.910 & (+0.806) & {\rm Beamline ~at ~Tevatron}\\
+0.918 & (+0.913) & {\rm Off-Diagonal ~at ~Tevatron}\\[0.2cm]
+0.305 & (+0.311) & {\rm Helicity ~at ~LHC(14 ~TeV)},
\end{array} \right. \nonumber
\end{eqnarray}
at LO with NLO in parenthesis, see   \cite{Bernreuther:2001rq}.
At the LHC, the coefficient $ C_{t\bar{t}}$ in the off-diagonal and beamline bases is small, $<0.10$.

\subsection{Azimuthal Correlations}
However, there are interference effects between the various spin components of the top-antitop system, e.g. between $t_L \bar{t}_L$ \& $t_R \bar{t}_R$ for like helicity gluon fusion, which leads to azimuthal correlations between the decay products,
\begin{eqnarray}
{1 \over \sigma_T}  {d\sigma \over d\Delta \phi}   = {1 \over 2} (1-D \cos \Delta \phi)
\end{eqnarray}
where in the ZM frame, the azimuthal correlations along the production axis are given by
\begin{eqnarray}
D=\left\{ \begin{array}{ll}
+0.132  & {\rm Tevatron}\\
-0.353 & {\rm LHC(14~TeV)}
\end{array} \right. 
\end{eqnarray}
from \cite{Bernreuther:2010ny}.
The azimuthal correlations about the beam axis in the Laboratory frame for the dilepton events are discussed in \cite{Mahlon:2010gw}, see Fig. 3.  These azimuthal correlations maybe easier to observe
than the other angular correlations.

\begin{figure}
\begin{center}
\includegraphics[scale=0.47]{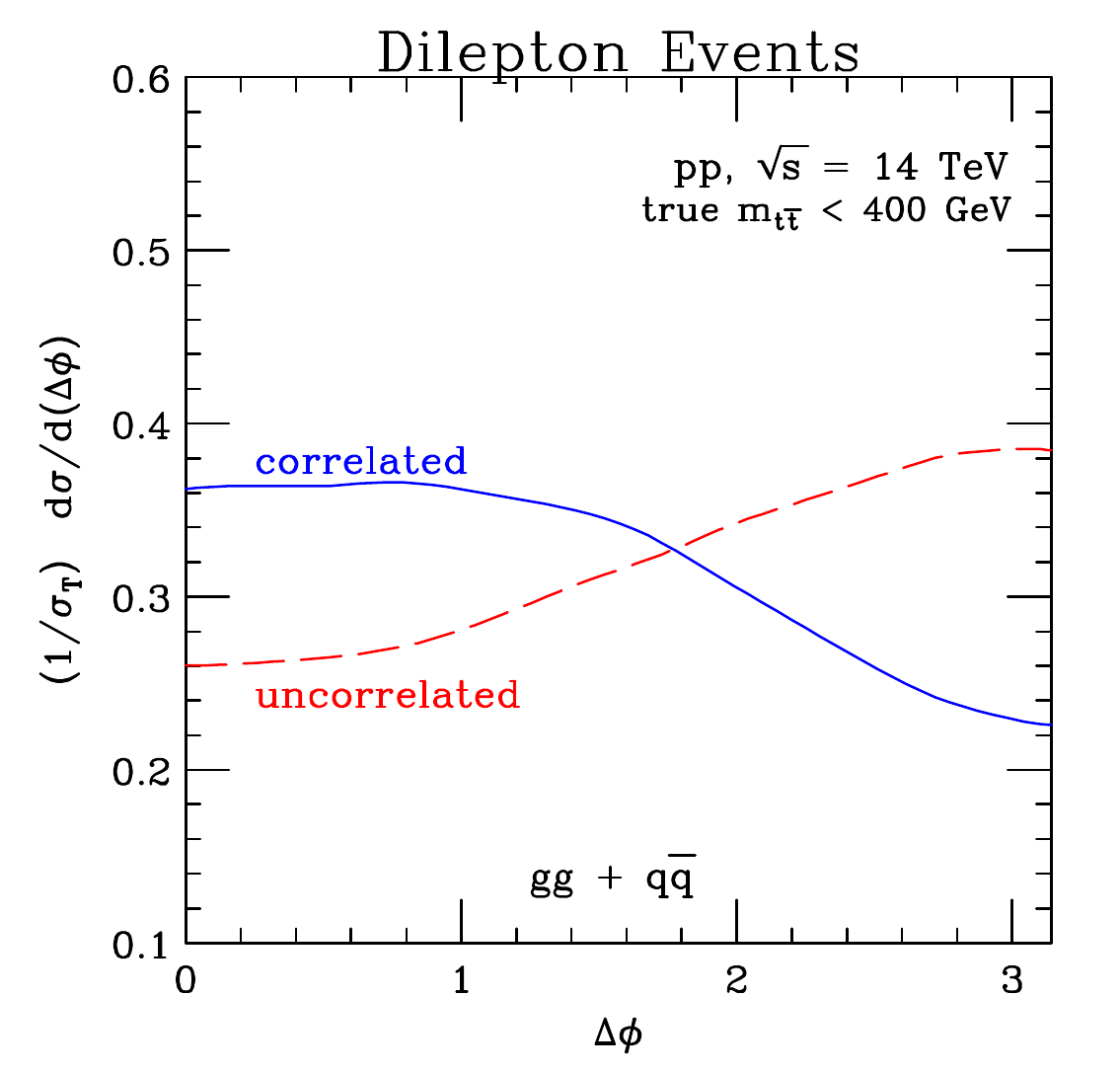}     
\end{center}
\caption{The $\Delta \Phi$ distribution of di-lepton events assuming the true $m_{t\bar{t}}<$ 400 GeV.}
\end{figure}

\acknowledgments
Fermilab is operated by the Fermi Research Alliance under contract 
no.~DE-AC02-07CH11359 with the U.S. Department of Energy.

\end{document}